\documentclass[sigconf]{acmart}
\AtBeginDocument{%
  }

\setcopyright{acmlicensed}
\copyrightyear{2026}
\acmYear{2026}
\acmDOI{XXXXXXX.XXXXXXX}
\acmConference[WWW '26]{The Web Conference 2026}{April 13--17,
  2026}{Dubai, UAE}
\acmISBN{978-1-4503-XXXX-X/2018/06}

\usepackage{tikz}
\usepackage{graphicx}

\usepackage{listings}
\usepackage{xcolor}

\usepackage{appendix}

\usepackage{booktabs}
\usepackage{multirow}

\usepackage{hyperref}
\hypersetup{hidelinks,
	colorlinks=true,
	allcolors=black,
	pdfstartview=Fit,
	breaklinks=true}

\usepackage{makecell}

\usepackage{setspace}
\usepackage{algorithmicx}

\usepackage{algpseudocode}
\usepackage{textcomp}
\usepackage{wrapfig}
\usepackage{textcomp}
\usepackage{pifont}
\usepackage{xcolor}
\usepackage{caption}
\usepackage{subcaption}
\usepackage{balance} 
\usepackage{xspace}
\usepackage[T1]{fontenc}
\usepackage[scaled=0.81]{beramono}
\usepackage{balance}

\usepackage{paralist, blindtext}

\usepackage{enumitem}  
\usepackage{url}
\usepackage[ruled,lined,linesnumbered,vlined]{algorithm2e}
\usepackage{multirow}
\usepackage{lscape}
\usepackage{longtable}
\usepackage{array} 
\usepackage{array}
\usepackage{tcolorbox}
\usepackage{color}
\usepackage[T1]{fontenc}
\usepackage[scaled=0.81]{beramono}
\usepackage{fancyvrb}
\usepackage{listings}
\usepackage{multicol}
\usepackage{lipsum}
\usepackage{booktabs}
\usepackage{listings}
\usepackage{color}
\usepackage{graphicx}
\usepackage{calc}
\usepackage{upgreek}
\usepackage{float}
\usepackage{threeparttable}
\usepackage[framemethod=tikz]{mdframed}
\usepackage{lipsum}
\usepackage{float}
\usepackage{pifont}

\usepackage{color}
\usepackage{colortbl}

\usepackage{float}

\newcommand{\tool}{OmniIntent}

\begin{document}

\title{\textsc{\tool}: A Trusted Intent-Centric Framework for User-Friendly Web3}


\author{Zhuoran Pan}
\orcid{0009-0001-8141-5587}
\affiliation{%
  \institution{Peking University}
  \city{Beijing}
  \country{China}
}
\email{vanthon256@gmail.com}

\author{Yue Li}
\affiliation{%
  \institution{Taiyuan University of Technology}
  \city{Taiyuan}
  \state{Shanxi}
  \country{China}
}
\email{liyue@tyut.edu.cn}

\author{Zhi Guan}
\affiliation{%
  \institution{Peking University}
  \city{Beijing}
  \country{China}
}
\email{guan@pku.edu.cn}

\author{Jianbin Hu}
\affiliation{%
  \institution{Peking University}
  \city{Beijing}
  \country{China}
}
\email{hujianbin@pku.edu.cn}

\author{Zhong Chen}
\affiliation{%
  \institution{Peking University}
  \city{Beijing}
  \country{China}
}
\email{zhongchen@pku.edu.cn}


\renewcommand{\shortauthors}{ et al.} 

\begin{abstract}
The increasingly complex Web3 ecosystem and decentralized finance (DeFi) landscape demand ever higher levels of technical expertise and financial literacy from participants. The Intent-Centric paradigm in DeFi has thus emerged in response, which allows users to focus on their trading intents rather than the underlying execution details. However, existing approaches, including Typed-intent design and LLM-driven solver, trade off expressiveness, trust, privacy, and composability.

We present \textsc{\tool}, a language–runtime co-design that reconciles these requirements. \textsc{\tool} introduces ICL, a domain-specific Intent-Centric Language for precise yet flexible specification of triggers, actions, and runtime constraints; a Trusted Execution Environment (TEE)-based compiler that compiles intents into signed, state-bound transactions inside an enclave; and an execution optimizer that constructs transaction dependency graphs for safe parallel batch submission and a mempool-aware feasibility checker that predicts execution outcomes. Our full-stack prototype processes diverse DeFi scenarios, achieving 89.6\% intent coverage, up to 7.3× throughput speedup via parallel execution, and feasibility-prediction accuracy up to 99.2\% with low latency.
\end{abstract}

\ccsdesc[500]{Do Not Use This Code~Generate the Correct Terms for Your Paper}
\ccsdesc[300]{Do Not Use This Code~Generate the Correct Terms for Your Paper}
\ccsdesc{Do Not Use This Code~Generate the Correct Terms for Your Paper}
\ccsdesc[100]{Do Not Use This Code~Generate the Correct Terms for Your Paper}

\keywords{Web3, Domain-Specific Language, Trusted Execution Environment, Intent-Centric}

\received{20 February 2007}
\received[revised]{12 March 2009}
\received[accepted]{5 June 2009}

\maketitle

\section{Introduction}
\label{Introduction}

\begin{figure}[h!]
    \centering
    \includegraphics[width=250bp]{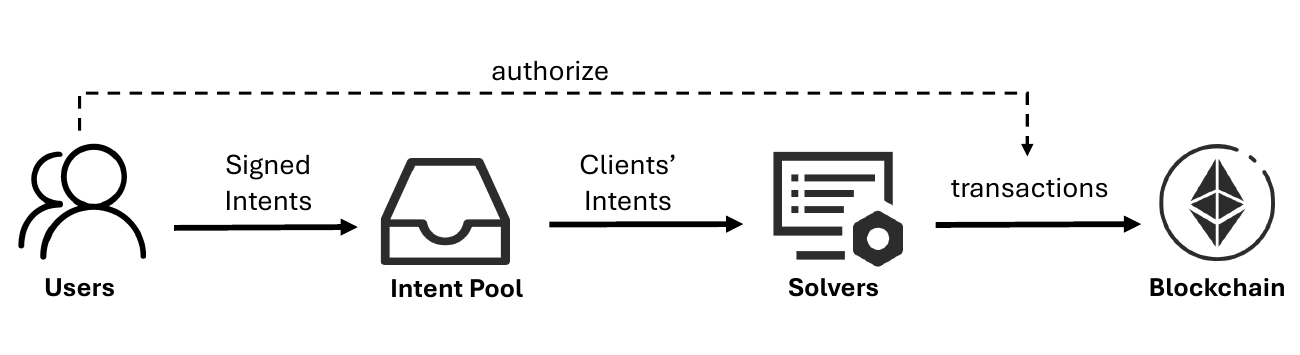}
    \caption{Typical workflow in intent-centric DeFi systems.}
    \label{intent_flow}
\end{figure}

Despite rapid technological advancements in Web3~\cite{wan2024web3}, end-user interaction remains largely \emph{transaction-centric}, \emph{i.e.}, users specify low-level contract calls, parameters, and fees~\cite{mohanta2018overview}. This forces them to learn protocol-specific interfaces across these DeFi (\emph{e.g.}, wallets, DEXs, lending markets, bridges), posing a significant barrier to apply traditional financial strategies in DeFi and confining Web3’s revolutionary potential to a niche audience~\cite{auer2024technology}. Recently, an \emph{intent-centric} paradigm has been proposed to revolutionize Web3 interaction: \emph{instead of specifying multiple low-level calls, users only need express what they want to achieve}~\cite{chitra2024analysis}. For example, a user can submit a single intent \emph{“stake 10 ETH, borrow USDC at rate $\le r$, purchase asset $X$, and bridge $X$ to BSC before $t$”} and the intent-centric system will compile and execute it without further users involvement, whereas a transaction-centric workflow would require at least four transactions from the user.
Figure \ref{intent_flow} shows the workflow of a typical intent-centric system: the signed intent will be submitted to the intent pool, the \emph{solver} automatically understands the intent and initiates necessary actions (\emph{i.e.}, transaction) to fulfill it by referring the DeFi interface and on-chain state, and then the \emph{solver} autonomously carries out these transactions on chain.
Note that a prior authorization from user is necessary to allow solver-issued transactions to act on user assets.

Currently, intent-centric designs advance along two directions. \emph{Typed-intent design}, such as UniswapX~\cite{bachu2025overview}, Cowswap~\cite{shi2024cows} and 1inch~\cite{onishchuk2024advancing}, support a limited types of intents (\emph{e.g.}, price-bounded swaps, cross-chain transfers) and execute them through competitive solvers; the design is efficient and easier to secure but does not generalize well to complex intents.
The second relies on \emph{LLM-driven solvers }that translates natural-language intents into transactions~\cite{nlok5923solver}, which increases expressiveness of intents while introducing opaque decision paths, weak interpretability, and added trust in both the model and its execution backend.

Overall, both directions have the following limitations: (i) an inherent trade-off between the expressiveness of intents and the accuracy with which they can be interpreted, \emph{i.e.}, typed-intent design methods support relatively simple user intents to achieve higher accuracy, whereas LLM-driven solvers allow natural-language expression of intents with inaccuracy of translation~\cite{huang2025dmind}; 
(ii) LLM-based solvers suffer from unverifiable and untrusted intent-to-transaction compilation, as their reasoning and generated transactions cannot be formally validated, while typed-intent systems remain verifiable but lack sufficient expressiveness for composite strategies; and
(iii) lack of support for composite intents, \emph{i.e.}, typed-intent design focuses on single-function intents such as token swaps, offering little support for intents that span multiple contracts or platforms. Moreover, LLM's reliability will be greatly reduced when handling multiple intents.

Proposing a solution that addresses these gaps entails the following challenges: ($C_1$) seek an appropriate intent expression method between the two extremes of typed design and LLM-based approaches, in order to achieve a delicate balance between intent expressiveness and interpretation accuracy; 
($C_2$) because the system involves the management of real assets, the intent-to-transaction parsing process must be trustworthy, and the parsed transactions must accurately implement the user’s diverse intent; and
($C_3$) for composite intents that may involve multiple interdependent transactions, their interaction and the potential risks of failure during execution need to be carefully considered. 

This paper overcomes these challenges with a trusted intent-centric framework. At its core is the Intent-Centric Language (ICL), a domain-specific language~\cite{mernik2005and} that captures user intents as structured programs composed of trigger conditions, transaction actions, and execution constraints. ICL supports both precise and fuzzy intents (e.g., “buy rare NFT” or “stake with low risk”), which supports diverse intents without sacrificing the accuracy of intent interpretation ($C_1$). It is worth noting that ICL is not intended for novice users to write directly. Instead, it serves as an intermediate representation that can be generated by UI tools or natural language front-ends, while remaining concise and auditable for advanced users.

To ensure that the system conversion process is reliable and accurate ($C_2$), we compile ICL inside a Trusted Execution Environment (TEE)~\cite{geppert2022trusted}, which preserves the confidentiality of users’ private keys and intent parameters, attests to the correctness of the compilation process, and produces signed transactions bound to real-time blockchain state. This TEE-based design enables pre-authorization and immediate execution under remote attestation, allowing intents to be executed securely and atomically even in constrained environments such as browsers or mobile devices.

To handle compilation and execution of multiple composite intents ($C_3$), we introduce an execution optimizer which includes a parallel transaction submitter that identifies opportunities for concurrent execution for multi-step intents while maintaining semantic correctness, and a feasibility checker that simulates multiple future blockchain states—including mempool~\cite{wang2024understanding} dynamics—to estimate execution success probabilities and provide risk feedback to users before submission.

We implement a full-stack prototype, \textsc{\tool}, integrating these three components into a cohesive system. \textsc{\tool} supports trusted intent compilation, transaction planning, dependency-aware parallel execution, and risk prediction.

Our contributions are as follows:

We conceptualize and formalize intent specification in DeFi as a programmatic abstraction. By designing ICL that allows for combination of conditions and expression of multiple intents, we show how structured yet flexible constructs can achieve high expressiveness without sacrificing semantic clarity or verifiability. To the best of our knowledge, we are the first to incorporate domain-specific language interaction paradigms into Web3 intent-centric architectures. Evaluation shows that ICL achieves high expressiveness (89.6\% of diverse DeFi intents).
    
We design and implement a trusted compiler that operates entirely inside a TEE. The compiler securely translates ICL intents into verifiable on-chain transactions while preserving the confidentiality of private keys and runtime parameters. It attests to both the integrity and correctness of the compilation process and produces signed transaction bundles bound to real-time blockchain states.

We introduce an execution optimizer that enables dependency-aware, risk-predictive intent execution. The optimizer constructs a transaction dependency graph to identify parallelizable components in composite intents and submits them concurrently while maintaining semantic correctness. A feasibility checker further simulates future blockchain states—including mempool dynamics—to estimate execution success probabilities and provide pre-execution risk feedback, achieving up to 7.3× throughput improvement and 99.2\% success prediction accuracy.

We build and open-source a full-stack prototype, \textsc{\tool}, integrating ICL, the trusted compiler, and the execution optimizer into a cohesive framework. The syntax design of ICL and the Java implementation of the resolution system can be found at \url{https://anonymous.4open.science/r/Intent-Centric-DeFi-DSL-3FF7}.

\section{Background}

\begin{figure*}[t]
    \centering
    \includegraphics[width=\textwidth]{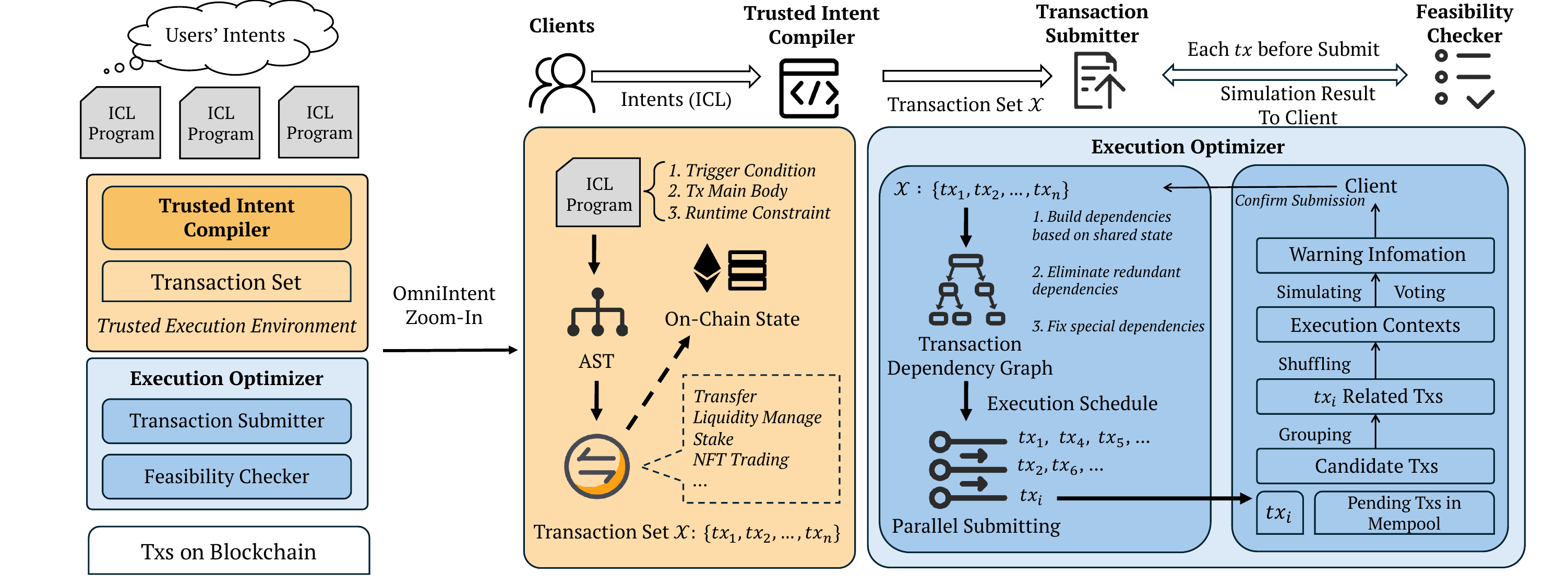}
    \caption{\textsc{\tool} Architecture.}
    \label{ICL-Resolver_architecture}
\end{figure*}

\subsection{Intent-Centric Framework}

Intent-centric frameworks replace users' direct submission of concrete transactions with declarative \textit{intents} that are posted to a shared \textit{intent pool} and consumed by a decentralized solver network~\cite{goes2023anoma,wang2024achieving,myakala2025intent}. 

Solvers scan the pool, match complementary intents, construct transaction bundles (often using matching, privacy-preserving proofs, and fee negotiation), and present concrete proposals which users then authorize (e.g., by signing) for on-chain submission. This separation of \textit{what} from \textit{how} improves liquidity aggregation and execution efficiency and lets users express richer goals without detailed market knowledge. Current systems, however, remain limited: advanced behaviors such as multi-step strategies, conditional logic, and tightly interdependent goals are  unsupported, and current frameworks lack formal semantics, a type system, or constraint-validation mechanisms necessary for safety and composability.

\subsection{Trusted Execution Environment}

Trusted Execution Environments (TEEs) provide hardware-enforced confidentiality and integrity for code and data running in isolated enclaves~\cite{jauernig2020trusted}. Representative implementations include Intel SGX, ARM TrustZone, and AMD SEV.~\cite{zheng2021survey,pinto2019demystifying,li2022systematic} 
Lightweight enclave OSes (e.g., Occlum, Eleos) and related abstractions reduce the trusted computing base and simplify secure system calls for complex workloads~\cite{shen2020occlum,orenbach2017eleos,paju2023sok,hardjono2019decentralized}. 

TEEs support hardware-backed remote attestation (signed measurements of code/configuration), in-enclave key generation, and sealed storage; together these features enable enclaves to obtain and protect long-term credentials, produce enclave-bound cryptographic proofs, and perform sensitive operations (e.g., signing or policy enforcement) without exposing secrets to the untrusted host, thereby creating a verifiable chain of trust from hardware to application. However, TEEs also introduce practical concerns such as vendor dependency and potential side-channel vulnerabilities, which should be considered when deploying production systems.

\section{Overview}
\label{ICL-Resolver Overview}

\textsc{\tool} is a comprehensive intent-centric framework that bridges the gap between user-friendly intent expression and secure, high-performance DeFi transaction execution.

\subsection{Architecture}

Figure~\ref{ICL-Resolver_architecture} illustrates the high-level architecture of \textsc{\tool}. The system takes client intents written in ICL as input and produces executable blockchain transactions as output. This system comprises three main parts.

\textbf{Intent Centric Language (ICL)} is a domain-specific declarative language for expressing user intents in DeFi. It provides syntactic constructs to describe actions such as asset swaps, conditional transfers, staking, liquidity provision, and other multi-step strategies. Each intent is represented as a structured statement composed of trigger conditions, transaction actions, and execution constraints. Trigger conditions define when an action should be initiated, transaction actions specify the operations to perform across protocols, and execution constraints capture limits such as slippage, gas usage, or deadlines. By separating these components, ICL enables a uniform and machine-readable representation of user strategies that can be compiled into executable transactions.
Unlike prior typed-intent specifications that only capture single-step actions, ICL is explicitly designed to express multi-step, composite, and dependency-aware strategies in a concise and structured manner, enabling complex DeFi workflows to be described as a single high-level program.

Taking clients' ICL programs as input, \textbf{Trusted Intent Compiler} operates inside a TEE and implements the parsing and translation pipeline for ICL programs. It uses an ANTLR-based~\cite{parr1995antlr} context-free grammar parser to convert ICL statements into Abstract Syntax Trees (ASTs). From the ASTs, the compiler extracts structured user intents composed of trigger conditions, transaction main body, and runtime constraints. Within the TEE, these structured intents are transformed into executable transactions by mapping high-level operations to corresponding smart contract interfaces of well-known platforms such as Uniswap, Sushiswap, Aave, Compound, etc. The compiler thus produces a set of well-formed transactions that correspond to the user’s specified actions while keeping all parsing and key material confined to the secure enclave.

Taking transaction set as input, \textbf{Execution Optimizer} integrates transaction submission and feasibility checking into a unified pipeline. It first applies a transaction dependency graph construction algorithm together with parallel scheduling optimization to maximize submission throughput while preserving dependency correctness. Before a transaction is submitted, it performs feasibility analysis by combining the transaction with a subset of mempool pending transactions~\cite{de2021analysis}, generating multiple execution contexts via randomized shuffling, simulating each context, and aggregating the results to assess execution risk. This integration ensures that submitted transactions are both efficiently scheduled and robust against execution failures.

\subsection{Illustrative Motivating Example}

\begin{figure}[htbp]
  \centering
\begin{lstlisting}[
  basicstyle=\ttfamily\scriptsize,
  keywordstyle=\color{blue},
  commentstyle=\color{green!50!black},
  stringstyle=\color{blue},
  numbers=left,               
  numberstyle=\scriptsize\color{gray},
  stepnumber=1,               
  numbersep=5pt,
  showstringspaces=false,
  tabsize=2,
  breaklines=true,
  breakatwhitespace=false,
  frame=single
]
swap 400000 USDC from wallet[0xA] for USDT on Uniswap checking slippage < 0.005;
swap 200000 USDC from wallet[0xA] for ETH on Uniswap checking slippage < 0.005 and fee < 150000;
buy popular price-increasing NFT using at most 80 ETH from wallet[0xA];
trigger balance wallet[0x_A] > 400000 USDT then    
    add 400000 USDC, 400000 USDT to Sushiswap receiving liquidity token to wallet[0xA];   
stake 160 ETH from wallet[0x_A] using long-term low-risk strategy checking price ETH > 4000; 
\end{lstlisting}
  \caption{Investment strategies written in ICL.}
  \label{investment_strategies}
\end{figure}

Consider a decentralized investment fund seeking to automate multi-protocol asset allocation and management through \textsc{\tool}. The authorized wallet initially holds a diversified portfolio consisting of 1 million USDC and 200 ETH. Its investment policy requires the following actions to be performed securely:

\emph{(i) Asset Rebalancing.} The user exchanges part of their USDC for USDT in order to join a high-yield stablecoin pool~\cite{reepu2025stablecoins}.

\emph{(ii) NFT Acquisition.} The user invests about 20\% of their portfolio in NFTs~\cite{wang2021non} with the potential for future appreciation.

\emph{(iii) Stablecoin Liquidity Provision.} Another 40\% of the portfolio is deposited into a stablecoin pool to earn trading fees and liquidity incentives.

\emph{(iv) Lending for Profit.} The remaining 40\% is supplied to lending protocols to earn long-term interest~\cite{castro2021financial}.

Using a non–intent-centric approach to achieve the above workflow requires users to understand token contracts, various staking protocols, NFT market dynamics, and risk management strategies.

Under existing intent-centric paradigms, for typed-intent designs, because they support only a very limited range of user intents, struggle to handle composite intents that span multiple contracts, such as NFT operations, staking, or liquidity-pool interactions. Conversely, although LLM-driven DApps can accept such intents, the uncertainty and low interpretability of the transactions generated by LLM planning strategies make this approach unacceptable in scenarios involving real assets and investments.

Using the ICL, however, the fund manager specifies the entire strategy declaratively as a set of intent statements shown in Figure~\ref{investment_strategies}. Specifically, line 1 to 2 rebalances assets by converting USDC into USDT and ETH; line 3 acquires promising NFTs with a capped ETH budget; line 4 to 5 uses a trigger condition to add both USDC and USDT to a Sushiswap liquidity pool once a balance threshold is reached; and line 6 stakes ETH under a low-risk strategy with a price constraint.

Upon submission, Trusted Intent Compiler parses the ICL program, extracts structured intents, calls related contracts to realize the intents, and constructs conditions and trading operations into a transaction set, which contains several tuples as Table~\ref{transaction_set} shows. 

\begin{table}
    \centering
    \caption{Transaction set which record trigger conditions, main trading operation, and runtime constraints.}
    \label{transaction_set}
    \resizebox{\columnwidth}{!}{
    \begin{tabular}{cccc}
        \toprule
         Idx & Trigger Condition & Main Operation & Runtime Constraint \\
        \midrule
         1 & / & uniswapRouter.exactInputSingle(...) & slippage $<$ 0.005 \\
        \midrule
         2 & / & uniswapRouter.exactInputSingle(...) & \makecell[l]{slippage $<$ 0.05 \&\& \\gasLimit $=$ 150000} \\
        \midrule
         3 & / & OpenSea.buyNFT(...) & / \\
        \midrule
         4 & balanceof(0x\_A) $>$ 400000 USDT & SushiSwapRouter.addLiquidity(...) & / \\
        \midrule
         5 & / & StakedAave.stake(...) & price ETH $>$ 4000 \\
        \bottomrule
    \end{tabular}
    }
\end{table}

\begin{figure}[h!]
    \centering
    \includegraphics[width=240bp]{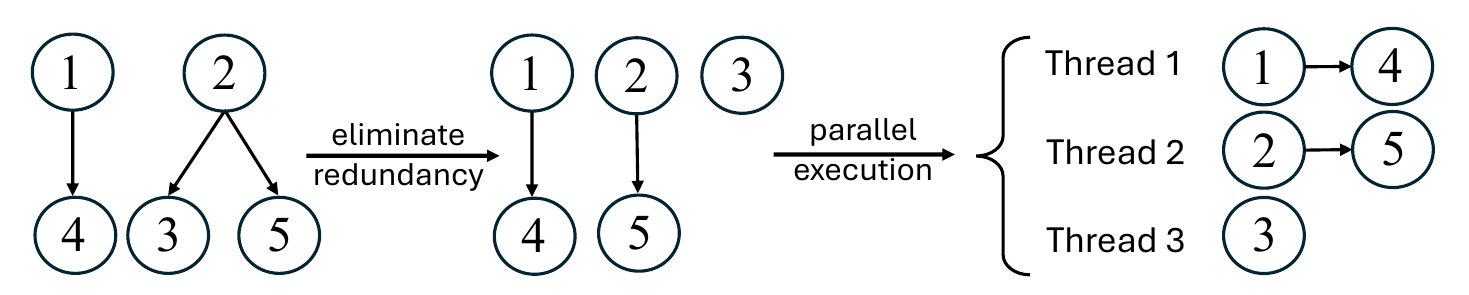}
    \caption{Transaction dependency construction and parallel submission execution.}
    \label{dependency_and_submission}
\end{figure}

Then the transaction set is submitted to the execution optimizer module as shown in Figure~\ref{dependency_and_submission}, which models the dependencies among transactions and eliminates redundant links. Based on the resulting dependency graph, together with each transaction’s trigger conditions and execution constraints, the system submits transactions to the blockchain in parallel. Before any transaction is dispatched on-chain, it is speculatively executed to anticipate potential execution risks in advance.

This motivating example highlights how \textsc{\tool} transforms complex, multi-step DeFi strategies into a single, auditable intent specification, allowing automated yet verifiable execution of diversified investment strategies, which cannot be accomplished by current intent-centric DApps.

\section{ICL Language Design}
\label{ICL Language Design}

The ICL is designed as a domain specific language that allows users to express DeFi intents in a programmable yet natural form. Its constructs are closely aligned with how users describe conditional transactions and asset management in decentralized finance, allowing for intuitive expression of triggers, execution logic, and runtime-checking conditions. The core syntax of ICL is shown in Figure~\ref{icl_rep_grammar}. The complete syntax of ICL is provided in the appendix~\ref{syntax_definition}.

\begin{figure}[htbp]
  \centering
    \begin{lstlisting}[basicstyle=\ttfamily\scriptsize,
keywordstyle=\color{blue},
commentstyle=\color{green!50!black},
stringstyle=\color{blue},
stepnumber=1,
numbersep=5pt,
showstringspaces=false,
tabsize=2,
breaklines=true,
breakatwhitespace=false,
language=python,
frame=single]
<program> ::= <triggerStatement> | <triggerStatement> <program> 
<triggerStatement> ::= ("trigger" condition "then")? statement ("checking" condition)? ";" 

<condition> ::= <andExpression> ("or" <andExpression>)*
<andExpression> ::= (<comparisonExpression> | <timeCondition>)("and" <comparisonExpression> | <timeCondition>))* 
<comparisonExpression> ::= <comparisonElement> <comparisonOperator> <comparisonElement> 
<comparisonElement> ::= <walletBalance> | <assetPrice> | <number> <asset> | <number> | "slippage" | "fee" | "(" orExpression ")" 

<statement> ::= <transferStatement> | <borrowStatement> | <repayBorrowStatement> | <swapStatement> | <addLiquidityStatement> | <removeLiquidityStatement> | <stakeStatement> | <buyNFTStatement> | <sellNFTStatement> 
    
    \end{lstlisting}
  \caption{Representative grammar of ICL.}
  \label{icl_rep_grammar}
\end{figure}

\begin{figure*}[t]
\centering
\scriptsize 
\begin{minipage}[t]{0.33\textwidth}
\begin{algorithmic}[1]
\State \textbf{function CompileICLProgram($P$):}
\State $txSet \gets \{ \}$
\State \textbf{For\ \ }{each $TS$ in $P$:} 
\State \quad $(T,S,C) \gets$ ParseTriggerAndStmt($TS$)
\State \quad $tx \gets$ CompileStatement($S$)
\State \quad $txSet.append(T,tx,C)$
\State \Return $txSet$

\Statex
\State \textbf{function CompileTranS($TranS$):}
\State $(sender,receiver,asset,amt)\gets$ Parse($TranS$)
\State $tx\gets$ERC20.transfer(sender,receiver,asset,amt)
\State\Return $tx$

\Statex
\State \textbf{function CompileSwapS($SwapS$):}
\State $(acct,fromAsset,toAsset,amt)\gets$ Parse($SwapS$)
\State $tx\gets$ Dex.exactInputSingle(acct,fromAsset,toAsset,amt)
\State\Return $tx$
\end{algorithmic}
\end{minipage}\hfill
\begin{minipage}[t]{0.33\textwidth}
\begin{algorithmic}[1]
\State \textbf{function CompileBorrS($BorrS$):}
\State $(acct,asset,platform,amt)\gets$ Parse($BorrS$)
\State $tx\gets$ platform.borrow(acct,asset,amt)
\State\Return $tx$

\Statex
\State \textbf{function CompileRepaS($RepaS$):}
\State $(acct,asset,platform,amt)\gets$ Parse($RepaS$)
\State $tx\gets$ platform.repayBorrow(acct,asset,amt)
\State\Return $tx$

\Statex
\State \textbf{function CompileAddlS($AddlS$):}
\State $(acct,assetPair,amt)\gets$ Parse($AddlS$)
\State $tx\gets$ Dex.mintLiquidity(acct,assetPair,amt)
\State\Return $tx$

\Statex
\State \textbf{function CompileRmvlS($RmvlS$):}
\State $(acct,assetPair,amt)\gets$ Parse($RmvlS$)
\State $tx\gets$ Dex.decreaseLiquidity(acct,assetPair,amt)
\State\Return $tx$
\end{algorithmic}
\end{minipage}\hfill
\begin{minipage}[t]{0.33\textwidth}
\begin{algorithmic}[1]

\State \textbf{function CompileStakS($StakS$):}
\State $(acct,asset,strategy)\gets$ Parse($StakS$)
\State info$\gets$QueryStakingPools(asset)
\State target$\gets$D.DecideStake(info,strategy)
\State $tx\gets$ StakeModule.stake(acct,asset,target)
\State\Return $tx$

\Statex
\State \textbf{function CompileBuynS($BuynS$):}
\State $(acct, budget, tend \gets)$ Parse($BuynS$)
\State info$\gets$QueryNFTMarket(...)
\State target$\gets$D.DecideNFT(info,budget,tend)
\State $tx\gets$OpenseaAPI.buildTx(acct,budget,target)
\State\Return $tx$

\Statex
\State \textbf{function CompileSelnS($SelnS$):}
\State $(acct, nft, tend \gets)$ Parse($SelnS$)
\State info$\gets$QueryNFTMarket(nft)
\State price$\gets$D.DecidePrice(info,tend)
\State $tx\gets$OpenseaAPI.buildTx(acct,nft,price)
\State\Return $tx$
\end{algorithmic}
\end{minipage}

\caption{ICL program compilation details ($P$: ICL program from client. $TS$: triggerStatement instruction in $P$. $T$: trigger condition. $S$: transaction statement. $C$: execution constraints. $TranS$: transfer statement. $BorrS$: borrow statement. $RepaS$: repay statement. $SwapS$: swap statement. $AddlS$: add liquidity statement. $RmvlS$: remove liquidity statement. $StakS$: stake statement. $BuynS$: buy NFT statement. $SelnS$: sell NFT statement. $tx$: transaction. $D$: decision module.)}
\label{compilation_alg}
\end{figure*}

\textbf{Trigger Statement.}  
The fundamental unit of an ICL program is the trigger statement. A trigger statement consists of three components:  
(i) an optional \emph{trigger condition} introduced by the keyword \texttt{trigger}, which specifies the on-chain state or temporal condition that must be met before execution;  
(ii) a mandatory \emph{transaction statement}, which encodes the intended blockchain operation;  
(iii) an optional \emph{runtime constraint} introduced by \texttt{checking}, which enforces execution-time restrictions such as slippage tolerance or gas limits.  

\textbf{Condition Expression.}  
Conditions appear both in trigger condition and runtime constraint. The grammar supports Boolean logic with \texttt{and}/\texttt{or}, enabling composite conditions. Supported predicates include wallet balance comparisons, asset price thresholds, slippage bounds~\cite{chemaya2024power}, fee limits~\cite{meister2024gas}, and time-based triggers. Nested expressions allow users to combine multiple state variables.

\textbf{Transaction Statement.}  
Transaction statements describe the actual operations to be executed once conditions are met. ICL supports a comprehensive set of DeFi-native operations, including:  
- \emph{Transfers}: token transfers;  
- \emph{Swaps}: asset exchanges on AMMs such as Uniswap or SushiSwap;  
- \emph{Borrowing/Lending}: interactions with lending protocols like Aave or Compound;  
- \emph{Liquidity Provision/Removal}: adding or removing liquidity in AMM pools;  
- \emph{Staking}: depositing assets into staking contracts;  
- \emph{NFT Trading}: purchasing or selling NFTs on supported marketplaces.  

\textbf{Execution Semantics.}  
The executability of an ICL program depends on the live blockchain state at compilation time. A program may be compiled only if its conditions are satisfiable and corresponding contract operations are available. Once compiled, each transaction statement is mapped to the respective protocol contracts (e.g., \texttt{UniswapRouter.exactInputSingle}, \texttt{AaveLendingPool.borrow}). This ensures that high-level intents are concretely realized as sequences of transactions consistent with on-chain semantics.

\section{Trusted Intent Compiler}
\label{Trusted Intent Compiler}

Taking ICL programs as input, Trusted Intent Compiler performs two main tasks: (1) checking the correctness of ICL programs; and (2) trustworthily compiling ICL instructions into a set of transactions.

\subsection{ICL Program Compilation}

ICL compilation is inherently state-dependent: each statement’s validity and the resulting transaction set hinge on real-time blockchain conditions such as asset prices, liquidity, and NFT market activity. To handle this, the compiler integrates on-chain data interfaces to fetch up-to-date information and interpret conditional expressions or fuzzy constraints in user intents. Figure~\ref{compilation_alg} presents details of ICL program compilation.

For a given program $P$, the compiler iterates over each trigger statement ($TS$), performing syntax analysis and decomposing it into a trigger condition ($T$), a transaction statement ($S$), and runtime constraints ($C$). The conditions in $T$ and $C$ are parsed and stored as data structures that specify, respectively, the execution trigger and runtime constraints for the transaction compiled from $S$. These constraints are enforced at runtime but only parsed during compilation.

The compiler then classifies $S$ and invokes the appropriate compilation routine. Transfer statements (\texttt{TranS}) map to ERC20 \texttt{transfer} calls. Borrow and repayment statements (\texttt{BorrS}, \texttt{RepaS}) compile to protocol-specific calls (e.g., \texttt{borrow}, \texttt{repayBorrow}). Swap statements (\texttt{SwapS}) invoke decentralized exchange functions such as \texttt{swap} or \texttt{exactInputSingle}. Liquidity management statements (\texttt{AddlS}, \texttt{RmvlS}) compile to \texttt{mint},  \texttt{decreaseLiquidity} operations.

More complex intents trigger decision-making modules. Staking statements (\texttt{StakS}) query platform data such as pool depth and APY, then call a decision module $D$ (e.g., weighted ranking, random forest, or LLM-based scoring) to select targets before emitting stake transactions. NFT buy/sell statements (\texttt{BuynS}, \texttt{SelnS}) gather market metrics (volume, price trends, holdings) and use $D$ to pick purchase or sale parameters before invoking the Opensea~\cite{white2022characterizing} API to build transactions.

Specifically, consider the ICL statement in line 3 of Figure~\ref{investment_strategies}.
During compilation, this instruction is parsed into a transaction statement specifying the NFT asset, the spending limit of 80 ETH, and the source wallet. The compiler then queries NFT marketplace data and decides to pick an NFT that meets the user's preferences, and constructs a purchase transaction that calls the appropriate marketplace contract or API (e.g., OpenSea’s \texttt{CreateItemOffer}) with the correct parameters.

Similarly, the statement in line 4 to 5 of Figure~\ref{investment_strategies} is parsed into a trigger condition (checking the USDT balance of wallet) and a liquidity-provision transaction. The compiler translates this into a sequence of ERC20 \texttt{approve} calls followed by a SushiSwap router \texttt{addLiquidity} call, ensuring that the liquidity tokens are credited to the specified wallet once the trigger condition is satisfied.

Once all statements are processed, the compiler emits a structured transaction set ready for execution, while preserving the ability to adapt at runtime when on-chain state changes.

\subsection{Intent Compiler in TEE}
We adapt the intent compiler to run inside a TEE.
We formalize the compilation of ICL programs into blockchain transactions as a secure multi-party protocol among three entities:

\begin{itemize}
    \item $\mathcal{U}$ (User): holds an ICL program $P \in \mathcal{L}_{\mathrm{ICL}}$ and approves assets to be traded to the externally owned account (EOA) controlled by TEE.
    \item $\mathcal{E}$ (Enclave): generates a key pair $k : <sk, pk>$ as the EOA and the Occlum LibOS~\cite{shen2020occlum} runs the compiler, which is attested to the user.
    \item $\mathcal{N}$ (Blockchain Node): an untrusted full node providing blockchain state snapshots (or market data) and corresponding proofs.
\end{itemize}

The blockchain is modeled as a state machine $\mathcal{B} = (S, T)$, where $S$ is the state space and $T : S \times \mathcal{X} \rightarrow S’$ is the transition function, with $\mathcal{X}$ the set of valid transactions.

The compiler is a mapping:
\[
\mathcal{C} : \mathcal{L}_{\mathrm{ICL}} \times S \rightarrow \mathcal{X}
\]
Given $(P, S)$, the compiler outputs a signed transaction sequence $\mathcal{X} = \langle tx_1, \dots, tx_m \rangle$ satisfying:

Semantic fidelity: $\mathcal{X}$ implements the operational semantics of $P$ with respect to $S$.

Authorization: each $tx_i$ is signed under $k$.

Non-leakage: $P, sk, S$ never appear in plaintext outside $\mathcal{E}$.

\paragraph{Protocol Overview.}

\begin{figure}[h!]
    \centering
    \includegraphics[width=260bp]{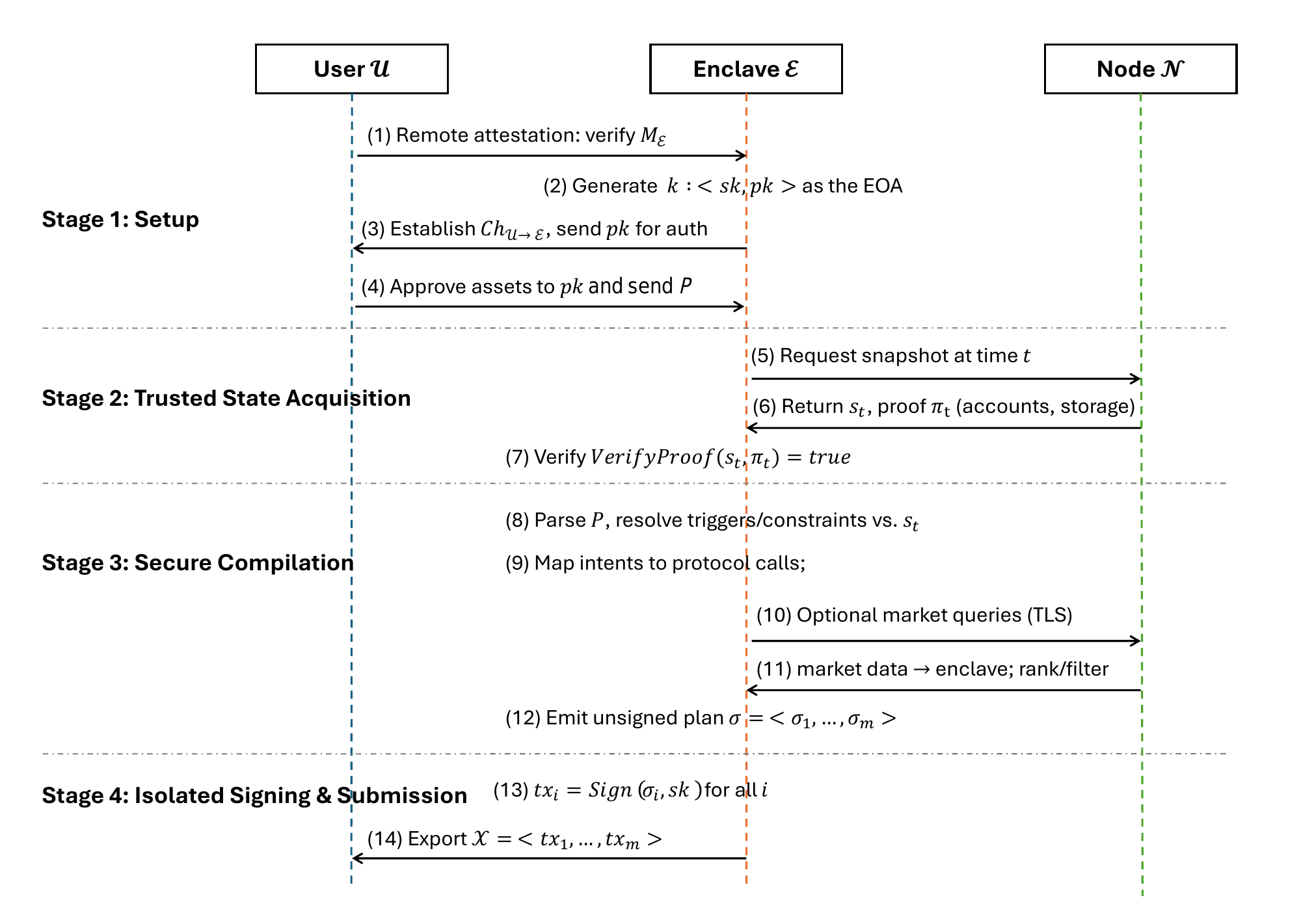}
    \caption{ICL compilation protocol.}
    \label{ICL_compilation_protocol}
\end{figure}

As shown in Figure~\ref{ICL_compilation_protocol}, the protocol proceeds in four stages, over secure channels $Ch_{\mathcal{U} \to \mathcal{E}}$ (user $\rightarrow$ enclave) and $Ch_{\mathcal{E} \leftrightarrow \mathcal{N}}$ (enclave $\leftrightarrow$ node):

\noindent
\textbf{Stage 1: Setup.}
\begin{enumerate}
    \item[$(1)$] $\mathcal{U} \to \mathcal{E}$: Initiate remote attestation to verify enclave measurement $M_{\mathcal{E}}$.
    \item[$(2)$] $\mathcal{E}$ generates key $<sk, pk>$ as the EOA.
    \item[$(3)$] Establish $Ch_{\mathcal{U} \to \mathcal{E}}$ and transmit $pk$ to $\mathcal{U}$.
    \item[$(4)$] $\mathcal{U}$ approves assets to be traded to the EOA $pk$ controlled by TEE.
\end{enumerate}

\noindent
\textbf{Stage 2: Trusted State Acquisition.}
\begin{enumerate}
    \item[$(5)$] $\mathcal{E} \to \mathcal{N}$: Request state snapshot at logical time $t$.
    \item[$(6)$] $\mathcal{N} \to \mathcal{E}$: Return $s_t$ and proof-of-state $\pi_t$ (e.g., Merkle proof~\cite{mizrahi2021optimizing} of account balances and contract storage).
    \item[$(7)$] $\mathcal{E}$ verifies $\mathrm{VerifyProof}(s_t, \pi_t) = \mathrm{true}$.
\end{enumerate}

\noindent
\textbf{Stage 3: Secure Compilation.}
\begin{enumerate}
    \item[$(8)$] $\mathcal{E}$ parses $P$ according to the ICL grammar, resolving triggers and constraints against $s_t$.
    \item[$(9)$] For basic DeFi intents, map to protocol-specific calls (e.g., ERC20 \texttt{transfer}, DEX \texttt{exactInputSingle}, lending protocol \texttt{borrow}), verifying feasibility within $\mathcal{E}$.
    \item[$(10)$] For complex DeFi intents, perform internal decision-making. External market data queries are performed over $Ch_{\mathcal{E} \leftrightarrow \mathcal{N}}$ or other TLS channels.
    \item[$(11)$] Decision module in $\mathcal{E}$ filters and ranks to seek solutions that meet user preferences.
    \item[$(12)$] Output unsigned transaction plan $\sigma = \langle \sigma_1, \dots, \sigma_m \rangle$.
\end{enumerate}

\noindent
\textbf{Stage 4: Isolated Signing.}
\begin{enumerate}
    \item[$(13)$] For each $\sigma_i \in \sigma$, compute $x_i = \mathrm{Sign}(\sigma_i, sk)$ inside $\mathcal{E}$.
    \item[$(14)$] Export $\mathcal{X} = \langle tx_1, \dots, tx_m \rangle$ to user for authorizing submission of transactions to the blockchain.
\end{enumerate}

\section{Execution Optimizer}
\label{Execution Optimizer}

Taking the transaction set as input, the execution optimizer submits transactions to Ethereum and predicts the risk of failure before each transaction is submitted, which consists of two parts: a parallel transaction submitter and a feasibility checker.

\begin{algorithm}[htbp]
	\footnotesize
	\DontPrintSemicolon
	\SetKwInOut{Input}{Input}
	\SetKwInOut{Output}{Output}
    \SetKw{Continue}{continue}
    \SetKwFunction{FSubmit}{processNode} 
    \SetKwProg{Fn}{Function}{}{end}             
	
	\Input{
		$\mathcal{X} \gets \{ tx_1, tx_2, \dots, tx_n \}$ is the transaction set.
	}
	
	\Output{
		$R$ is the report on executed transactions.
	}
	
	$WA_{in}, WA_{de} \gets \{\}$ \tcp*[r]{Track wallet inflows/outflows}
	$G \gets$ initialize dependency graph \newline \\
	
	\tcp{Phase 1: Construct primary dependencies.}
	$\mathcal{X}_{sorted} \gets$ sortByIntentOrder($\mathcal{X}$)
	
	\For{$tx \in \mathcal{X}_{sorted}$}{
		$asset_{in}, wallet_{in} \gets x$.getIncrease() \;
		$asset_{de}, wallet_{de} \gets x$.getDecrease() \;
		$G$.addVertex($tx$) \;
		\For{$tx_{pa} \in WA_{in}.get(\langle asset_{de}, wallet_{de} \rangle)$}{
			$G$.addEdge($tx_{pa}$, $tx$)
		}
		$WA_{in}$.put($\langle asset_{in}, wallet_{in} \rangle$, $tx$)\;
		$WA_{de}$.put($\langle asset_{de}, wallet_{de} \rangle$, $tx$)\;
	}
	
	\tcp{Phase 2: Simplify dependency graph.}
	$\mathcal{X}_{topo} \gets G$.topologicalSort().reverse()
	
	\For{$tx \in \mathcal{X}_{topo}$}{
		$F_{abort} \gets$ multipleKnapsack($tx$.getParents(), $tx$.getPredictedBalance()$)$\;
		adjustEdgesAndBalances($tx, F_{abort}, G$) \tcp*[r]{Rewire and adjust}
	}

	\tcp{Phase 3: Concurrent validation and execution.}
	initializeExecutionContext($G$) \tcp*[r]{Counters, flags, and thread pool}
	
	executeGraphConcurrently($G$, \texttt{processNode}) \tcp*[r]{Submit nodes with all parents done}
	
	\Fn{\texttt{processNode}($node$)}{
		\If{isSkipped($node$)}{
			$R$.log(SKIPPED)
		}\Else{
			checkFeasibilityAndWaitForUserConfirm($node$)\;
			$result \gets$ submitTransaction($node$)\;
			$R$.log($result$)
			
			\If{$result \in \{$SKIPPED, FAILED$\}$}{
				propagateSkipFlag($node$)
			}
		}
		notifySuccessors($node$) \tcp*[r]{Trigger child nodes if ready}
	}
	
	\textbf{Wait} until all nodes are processed.\\[3pt]
	\Return{$R$}
	
	\caption{Optimally Execute Transaction Sets}
	\label{optimally_execute_tx_set}
\end{algorithm}

\subsection{Parallel Transaction Submitter}
\label{Parallel Transaction Submitter}

ICL programs often contain multiple intents that map to transactions sharing states (accounts or tokens), creating implicit dependencies. Executing strictly in user-written order~\cite{wan2019evaluating} is inefficient. As shown in phase 1 of Algorithm~\ref{optimally_execute_tx_set}, we therefore build a \emph{Transaction Dependency Graph} (TDG), in which nodes are transactions and edges denote either asset-flow or protocol-level constraints (e.g., collateral must precede borrowing).

To increase concurrency, the optimizer prunes redundant edges while ensuring no unsafe asset underflows occur, as shown in phase 2 of Algorithm~\ref{optimally_execute_tx_set}. This pruning procedure is formalized as an optimization problem (see appendix~\ref{appendix:optimizer_theory}) and solved using a dynamic programming–based algorithm shown in appendix~\ref{multiple_knapsack}. Figure~\ref{update_the_DAG} shows how the TDG is updated after some dependencies of $N_1$ are removed (line 15 in Algorithm~\ref{optimally_execute_tx_set}).

\begin{figure}[h!]
    \centering
    \includegraphics[width=250bp]{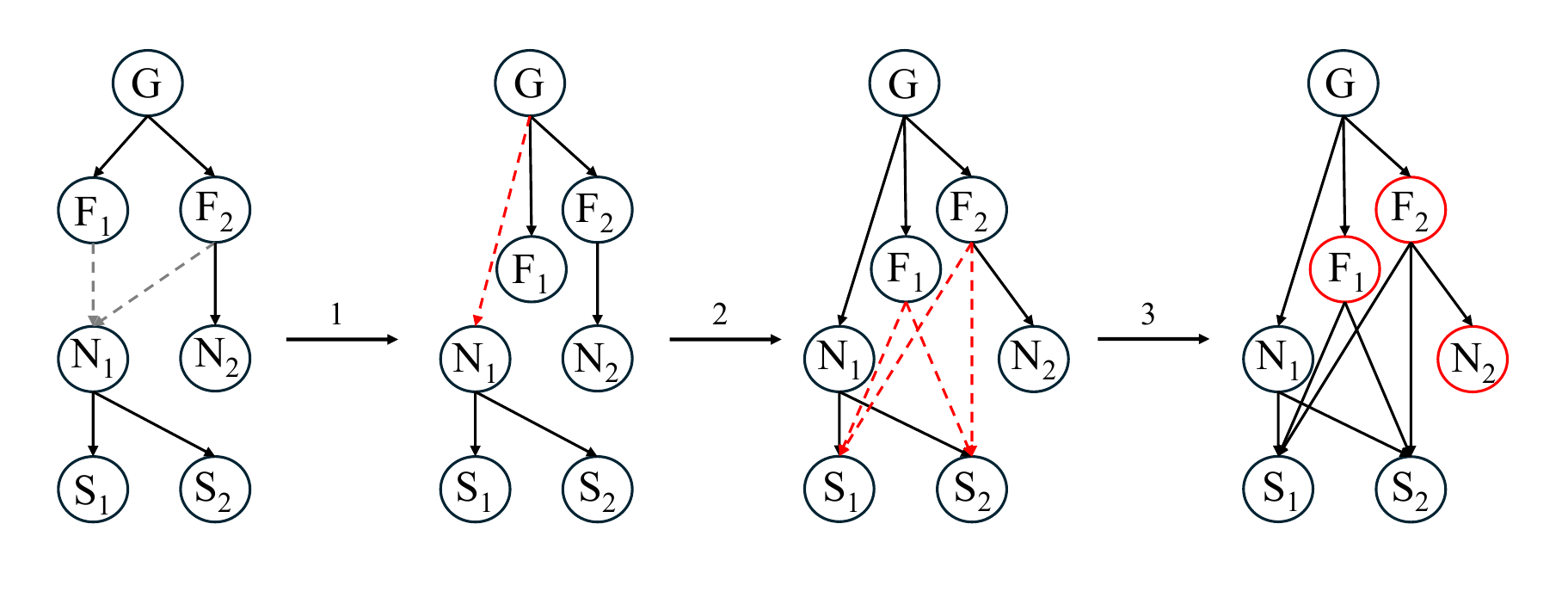}
    \caption{Remove dependency of $N_1$ and update the TDG}
    \par\vspace{0.5ex}
    \begin{minipage}{0.9\linewidth}
    \footnotesize 
    1) Add dependency edges from $N_1$ to its grandparent nodes. 
    2) Fix the dependencies of $N_1$'s child nodes on removed parent nodes. 
    3) Maintain the predicted balance of the removed parent nodes and their descendants.
    \end{minipage}
    \label{update_the_DAG}
\end{figure}

Protocol-specific dependencies that are not derivable from balances (e.g., collateralization rules) are captured via domain-specific patterns.  

Given the optimized TDG, a trigger-based engine submits ready transactions in parallel shown in phase 3 of Algorithm~\ref{optimally_execute_tx_set}. Failed transactions prevent unsafe execution of successors, ensuring robustness. This design achieves both correctness and responsiveness. Practical scalability considerations are deferred to appendix~\ref{appendix:optimizer_theory}.

\subsection{Feasibility Checker}
\label{Intent Feasibility Checker}

\begin{figure}[h!]
    \centering
    \includegraphics[width=250bp]{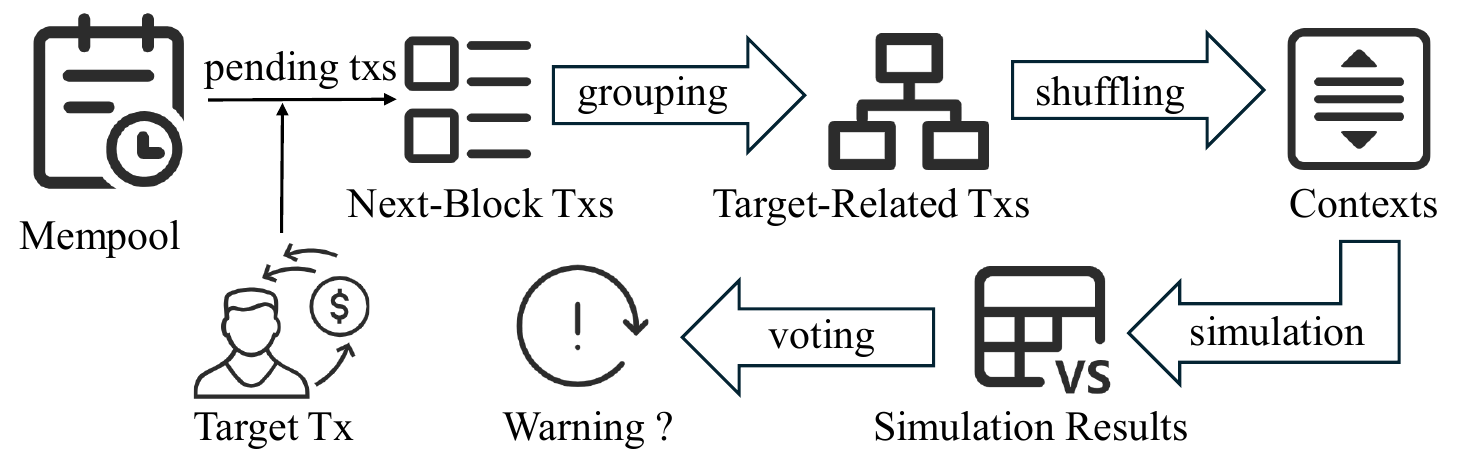}
    \caption{Feasibility checking framework.}
    \label{feasibility_checking_framework}
\end{figure}

Applied in phase 3 of algorithm~\ref{optimally_execute_tx_set}, the feasibility checker determines whether an intent-generated transaction can safely execute under both user-defined constraints and current blockchain conditions. Its core is a simulation-based analysis that estimates transaction success probability in near-future states.

As shown in figure~\ref{feasibility_checking_framework}, the checker retrieves pending transactions from the mempool via RPC, sorts them by gas price to approximate miner priorities, and aggregates them up to the block gas limit to form a predicted next block. It then constructs execution contexts for the target transaction by grouping related transactions using a disjoint-set algorithm and generating randomized variants to capture execution variability.

Each context is simulated up to the target transaction, and success rates across contexts determine risk level: transactions succeeding in most simulations are executed, while others trigger warnings. This simulation-based feasibility check extends beyond static validation, providing proactive risk assessment under dynamic mempool and state changes.

\section{Evaluation}
\label{Evaluation}

We evaluate system performance from three aspects: expressiveness of ICL, efficiency of the compiler, and accuracy of the feasibility checker. Testbed and datasets are defined below.

\textbf{Testbed.} Our system is deployed on a server featuring an Intel i7 eight‑core processor, 16 GB of RAM, and a 1 Gbps Ethernet Internet connection. The operating system on the host is Ubuntu 20.04 LTS. Within the trusted execution environment, Intel SGX SDK (v2.25)~\cite{tian2019practical} is employed, and Occlum (v0.30.1)~\cite{shen2020occlum} serves as the lightweight enclave OS for running Java JAR applications. For blockchain integration, both the Sepolia public test network~\cite{rani2023academic} and a customizable private Ethereum chain, operated locally via Geth~\cite{rouhani2017performance}, are utilized.

\textbf{Dataset 1.} We use the DeFi user transaction intent dataset constructed in the work of Bob the Solver~\cite{nlok5923solver} as the initial test set. Due to the limited availability of similar datasets in existing research to the best of our knowledge, we construct a comprehensive dataset of user transaction intents by applying mutation and augmentation techniques using a large language model (GPT-4o).

\textbf{Dataset 2.} We construct a synthetic dataset using a program generator that produces ICL files of varying sizes. The generator randomly samples different categories of intents and instantiates instruction parameters with on-chain information as well as the contextual dependencies among previously generated instructions, thereby ensuring the validity of the resulting ICL commands.

\textbf{Dataset 3.} We construct a dataset of high-frequency transaction flows to simulate realistic blockchain workloads. The flows are generated by JavaScript scripts that continuously submit transactions including transfer, swap, and liquidity management to a locally deployed Ethereum network, reproducing the congested execution conditions observed in practice.

\subsection{Expressiveness of ICL}

\paragraph{Intent Expressiveness.}
We prompt GPT-4o to learn ICL grammar (appendix~\ref{prompts}) and sequentially input intents from dataset~1. Outputs are parsed by our semantic module to verify conformance. Coverage is calculated as the proportion of intents successfully mapped into valid ICL instructions (results in Table~\ref{ICL_coverage}).

ICL achieves high coverage for DeFi intents: \emph{Borrow} and \emph{Repay} reach 97.8\% and 99.4\%; \emph{Swap}, \emph{Add Liquidity}, and \emph{Remove Liquidity} exceed 95\%. In contrast, NFT-related intents remain low (\emph{Purchase}: 59.6\%, \emph{Sale}: 47.9\%) due to heterogeneous market semantics. Overall, ICL covers 3629 of 4050 intents (89.6\%).

Although ICL covers most common DeFi primitives, its expressiveness is limited to the predefined operator set. Intents requiring arbitrary contract logic or unsupported cross-chain behaviors cannot be directly expressed and may require extending the operator library or falling back to lower-level interactions. We therefore position ICL as a practical, safe middle ground rather than a fully Turing-complete language.

\paragraph{Language Usability.}
We decompose ICL into triggers, actors, and checks, and compare to Solidity/Vyper implementations generated with GPT-4o (appendix~\ref{prompts}). Two metrics are used: relative code length (rl) and relative similarity to natural language (rs)~\cite{munk2022names} (Table~\ref{code_length_similarity}).

\begin{table}
    \centering
    \caption{ICL intent expressiveness coverage.}
    \label{ICL_coverage}
     \resizebox{\columnwidth}{!}{
    \begin{tabular}{lccr}
        \toprule
        Intent Type & Total Num & Success Num & Coverage \\
        \midrule
         Transfer & 480 & 391 & 81.5\% \\
         Borrow & 500 & 489 &  97.8\% \\
         Repay & 500 & 497 &  99.4\% \\
         Swap & 500 & 494 &  98.8\% \\
         Add Liquidity & 500 & 483 &  96.6\% \\
         Remove Liquidity & 500 & 498 &  99.6\% \\
         Stake & 580 & 513 &  88.4\% \\
         NFT Purchase & 250 & 149 &  59.6\% \\
         NFT Sale & 240 & 115 &  47.9\% \\
         \textbf{Total }& \textbf{4050} & \textbf{3629} & \textbf{89.6\%} \\
        \bottomrule
    \end{tabular}
    }
\end{table}

\begin{table}
    \centering
    \caption{Language usability.}
    \label{code_length_similarity}
    \resizebox{\columnwidth}{!}{
    \begin{tabular}{llcccc}
        \toprule
        Type & Action & Solidity(rl) & Vyper(rl) &  Solidity(rs) & Vyper(rs))\\
        \midrule
        \multirow{4}{*}{Trigger}
        & Asset Price & 12.44 & 6.037 & 0.146 & 0.178 \\
        & Slippage & 1.917 & 2.875 & 0.771 & 0.641  \\
        & Time & 4.256 & 1.897 & 0.349 & 0.405  \\
        & Balance & 2.575 & 1.466 & 0.251 & 0.483 \\
        \midrule
        \multirow{9}{*}{Main Tx} 
        & Transfer & 1.454 & 3.176 & 0.283 & 0.234 \\
        & Borrow & 5.472 & 6.042 & 0.183 & 0.141 \\
        & Repay & 8.042 & 3.915 & 0.215 & 0.333 \\
        & Swap & 3.671 & 8.949 & 0.124 & 0.058 \\
        & Add Liquidity & 6.527 & 8.391 & 0.191 & 0.193 \\
        & Remove Liquidity & 4.795 & 6.137 & 0.244 & 0.248 \\
        & Stake & 9.357 & 7.101 & 0.103 & 0.145 \\
        & NFT Purchase & 2.406 & 4.123 & 0.346 & 0.246 \\
        & NFT Sale & 3.993 & 3.392 & 0.240 & 0.234 \\
        \midrule
        \multirow{5}{*}{Checker} 
        & Asset Price & 16.83 & 10.96 & 0.098 & 0.274 \\
        & Slippage & 2.190 & 6.857 & 0.615 & 0.196 \\
        & Time & 2.639 & 1.889 & 0.257 & 0.380 \\
        & Balance & 2.157 & 1.241 & 0.368 & 0.597 \\
        & Tx Fee & 1.526 & 1.474 & 0.614 & 0.399 \\
        \bottomrule
    \end{tabular}
    }
\end{table}

Solidity code is often an order of magnitude longer: e.g., \emph{Asset Price} triggers require 12.44× more characters in Solidity, 6.04× in Vyper. Even \emph{Transfer} is 1.45× and 3.18× longer. ICL also preserves higher semantic similarity: \emph{Slippage} checks score 0.771 (Solidity) and 0.641 (Vyper) vs. ICL baseline, while complex ops like \emph{Stake} drop to 0.103 and 0.145. These results confirm ICL’s conciseness and semantic alignment.

\subsection{Efficiency of Compiler}

\begin{table}
    \centering
    \caption{Gas consumption of different transaction methods.}
    \label{ICL_gas}
    \resizebox{\columnwidth}{!}{
    \begin{tabular}{lcccc}
        \toprule
        Intent Type & Conventional & Cowswap & 1Inch Fusion & ICL \\
        \midrule
         Transfer & 21047 & 21471 & 22862 & 22176 \\
         Borrow & 283859 & / & / & 298082 \\
         Repay & 163013 & / & / & 157122 \\
         Swap & 126074 & 148050 & 108136 & 110793 \\
         Add Liquidity & 383796 & / & / & 346948 \\
         Remove Liquidity & 119692 & / & / & 147062 \\
         Stake & 191183 & / & / & 199840 \\
         NFT Purchase & 329755 & / & / & 332154 \\
         NFT Sale & 357423 & / & / & 341908 \\
        \bottomrule
    \end{tabular}
    }
\end{table}

\begin{figure}[h!]
    \centering
    \includegraphics[width=250bp]{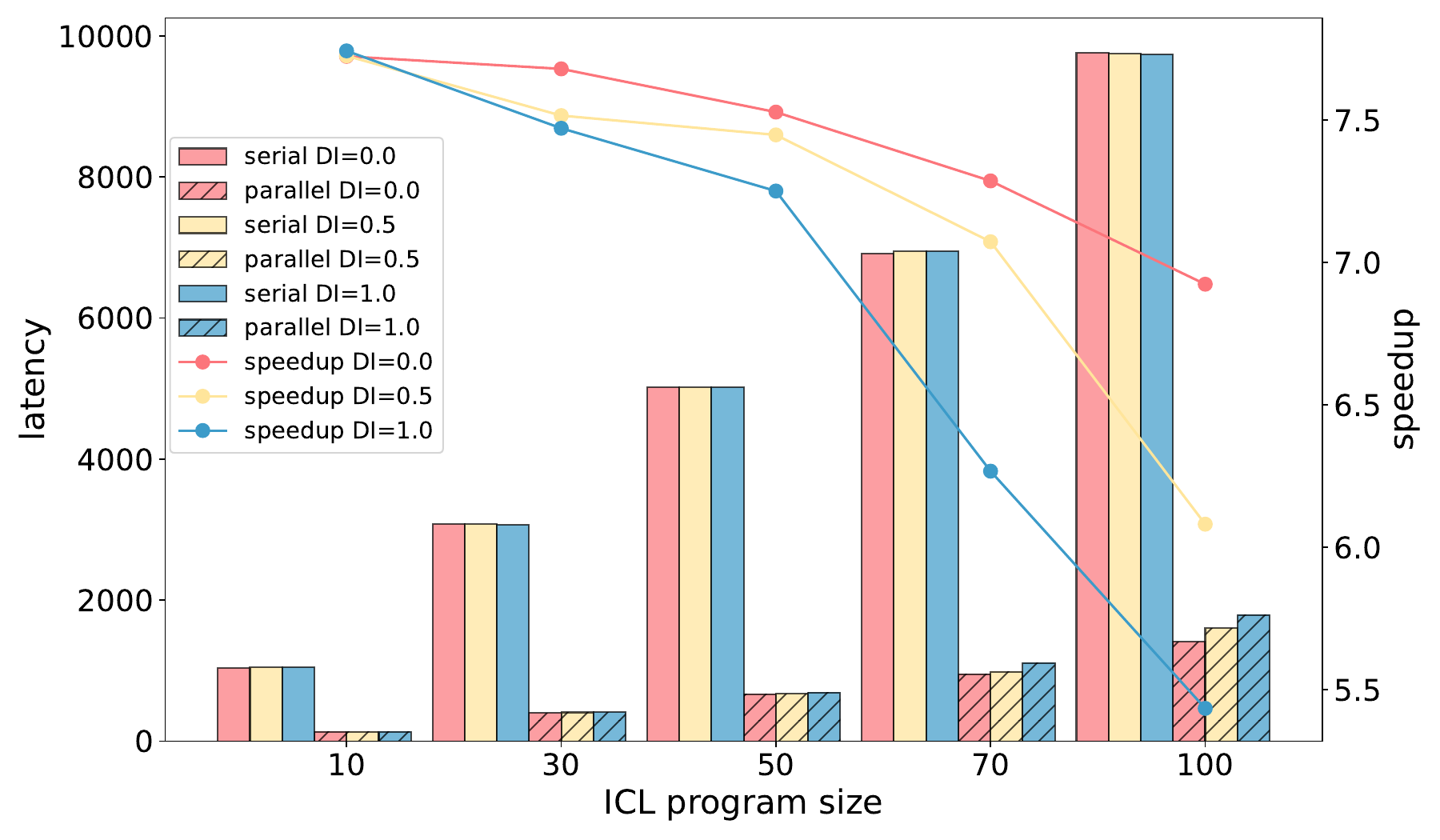}
    \caption{Latency and speedup of transaction sets under serial and parallel submission modes.}
    \label{trans_efficiency}
\end{figure}

\paragraph{Transaction Cost.}
Gas cost is measured on Sepolia using dataset~1 and compared with conventional transactions and protocols such as Cowswap and 1inch Fusion~\cite{kim2024optimal} (Table~\ref{ICL_gas}). 

ICL shows no systematic overhead: \emph{Transfer} and \emph{Swap} consume 22,176 and 110,793 gas, comparable to Cowswap (21,471, 148,050) and 1inch (22,862, 108,136). For complex ops, ICL is close to conventional: \emph{Borrow} (298,082 vs. 283,859), \emph{Repay} (157,122 vs. 163,013), and reduces cost in some cases (\emph{Add Liquidity}: 346,948 vs. 383,796). Results show general parity while offering broader expressiveness.

\paragraph{Transaction Efficiency.}

Using dataset~2, we evaluate throughput under serial vs. parallel execution. Dependency Index (DI\footnotemark) is varied at 0.0, 0.5, and 1.0, and latency is measured end-to-end. Results (Figure~\ref{trans_efficiency}) show consistent parallel speedups: at DI=1.0 with 50 txs, parallelism yields 7.3× acceleration; with 100 txs, 5.4×. Speedup declines with scale due to denser dependency graphs, but remains $>$5× even with 100 transactions, validating the dependency-graph optimization.

\footnotetext{%
Formally, the Dependency Index is defined as: 
\[
\text{DI}(x) =  P\!\left(\exists\, y < x \;\text{s.t.}\; \text{ICL}_x \text{ depends on } \text{ICL}_y \right)
\]
}

\subsection{Accuracy of Feasibility Checker}




\begin{table}
\centering
\caption{Prediction performance of the feasibility checker.}
\label{feasibility_checker_prediction}
\begin{tabular}{lccc}
\toprule
Method & Accuracy & Precision & Recall \\
\midrule
Naïve baseline (no mempool) & 0.836 & 0.971 & 0.845 \\
\midrule
OmniIntent ($K=1.0$) & \textbf{0.992} & \textbf{1.000} & \textbf{0.991} \\
OmniIntent ($K=1.5$) & 0.960 & 1.000 & 0.955 \\
OmniIntent ($K=2.0$) & 0.893 & 1.000 & 0.881 \\
\bottomrule
\end{tabular}
\end{table}

We evaluate the feasibility checker against a naïve baseline that predicts outcomes solely based on current on-chain states without mempool awareness. The baseline achieves only 83.6\% accuracy, 97.1\% precision, and 84.5\% recall.

As shown in Table~\ref{feasibility_checker_prediction}, our checker substantially improves prediction quality. With the default $K=1$, it achieves 99.2\% accuracy, 100\% precision, and 99.1\% recall, with no false positives. Increasing $K$ makes predictions more conservative but lowers recall. Prediction latency averages 225.8,ms per transaction (max $<$400,ms), introducing minimal overhead.

\section{Conclusion}

This work presents \textsc{\tool}, a trusted framework for intent-centric DeFi. It introduces ICL, a domain-specific language that expresses triggers, actions, and constraints for complex financial intents; a TEE-based compiler that attests code integrity, generates enclave-controlled EOAs, and signs state-bound transactions; and a dependency-graph optimizer with a mempool-aware feasibility checker enabling parallel and predictable execution. Together, these components achieve privacy-preserving, auditable, and high-throughput intent execution.



\bibliographystyle{ACM-Reference-Format}
\bibliography{sample-sigconf}

@String{Computing = "Computing" }

@String{Computer = "{IEEE} Computer" }

@String{Academic = "Academic Press" }

@String{Springer = "Springer-Verlag" }

@article{chitra2024analysis,
  title={An analysis of intent-based markets},
  author={Chitra, Tarun and Kulkarni, Kshitij and Pai, Mallesh and Diamandis, Theo},
  journal={arXiv preprint arXiv:2403.02525},
  year={2024}
}

@article{goes2023anoma,
  title={Anoma: a unified architecture for full-stack decentralised applications},
  author={Goes, Christopher and Yin, Awa Sun and Brink, Adrian},
  journal={Anoma Research Topics, Aug},
  year={2023}
}

@article{onishchuk2024advancing,
  title={Advancing DeFi Analytics: Efficiency Analysis with Decentralized Exchanges Comparison Service},
  author={Onishchuk, Evgenii and Dubovitskii, Maksim and Horch, Eduard},
  journal={arXiv preprint arXiv:2411.01950},
  year={2024}
}

@misc{nlok5923solver,
  title     = {Bob the Solver},
  author    = {Nitanshu Lokhande},
  url       = {https://github.com/nlok5923/solver},
  year      = {2023},
}

@article{parr1995antlr,
  title={ANTLR: A predicated-LL (k) parser generator},
  author={Parr, Terence J. and Quong, Russell W.},
  journal={Software: Practice and Experience},
  volume={25},
  number={7},
  pages={789--810},
  year={1995},
  publisher={Wiley Online Library}
}

@inproceedings{tian2019practical,
  title={A practical intel sgx setting for linux containers in the cloud},
  author={Tian, Dave and Choi, Joseph I and Hernandez, Grant and Traynor, Patrick and Butler, Kevin RB},
  booktitle={Proceedings of the Ninth ACM Conference on Data and Application Security and Privacy},
  pages={255--266},
  year={2019}
}

@inproceedings{shen2020occlum,
  title={Occlum: Secure and efficient multitasking inside a single enclave of intel sgx},
  author={Shen, Youren and Tian, Hongliang and Chen, Yu and Chen, Kang and Wang, Runji and Xu, Yi and Xia, Yubin and Yan, Shoumeng},
  booktitle={Proceedings of the Twenty-Fifth International Conference on Architectural Support for Programming Languages and Operating Systems},
  pages={955--970},
  year={2020}
}

@inproceedings{shi2024cows,
  title={From CoWs to Multi-Chain AMMs: A Strategic Optimization Model for Enhancing Solvers},
  author={Shi, Zeshun and Sweck, Sydney and Zaki, Omar},
  booktitle={2024 IEEE International Conference on Blockchain (Blockchain)},
  pages={97--104},
  year={2024},
  organization={IEEE}
}

@article{bachu2025overview,
  title={An overview of uniswap v4 for researchers},
  author={Bachu, Brad and Hasbrouck, Joel and Saleh, Fahad and Wan, Xin},
  journal={Available at SSRN 5152215},
  year={2025}
}

@article{wan2024web3,
  title={Web3: The next internet revolution},
  author={Wan, Shicheng and Lin, Hong and Gan, Wensheng and Chen, Jiahui and Yu, Philip S},
  journal={IEEE Internet of Things Journal},
  volume={11},
  number={21},
  pages={34811--34825},
  year={2024},
  publisher={IEEE}
}

@inproceedings{mohanta2018overview,
  title={An overview of smart contract and use cases in blockchain technology},
  author={Mohanta, Bhabendu Kumar and Panda, Soumyashree S and Jena, Debasish},
  booktitle={2018 9th international conference on computing, communication and networking technologies (ICCCNT)},
  pages={1--4},
  year={2018},
  organization={IEEE}
}

@article{auer2024technology,
  title={The technology of decentralized finance (DeFi)},
  author={Auer, Raphael and Haslhofer, Bernhard and Kitzler, Stefan and Saggese, Pietro and Victor, Friedhelm},
  journal={Digital Finance},
  volume={6},
  number={1},
  pages={55--95},
  year={2024},
  publisher={Springer}
}

@article{huang2025dmind,
  title={DMind Benchmark: Toward a Holistic Assessment of LLM Capabilities across the Web3 Domain},
  author={Huang, Enhao and Sun, Pengyu and Lin, Zixin and Chen, Alex and Ouyang, Joey and Wang, Hobert and Dong, Dong and Zhao, Gang and Yi, James and Li, Frank and others},
  journal={arXiv preprint arXiv:2504.16116},
  year={2025}
}

@article{mernik2005and,
  title={When and how to develop domain-specific languages},
  author={Mernik, Marjan and Heering, Jan and Sloane, Anthony M},
  journal={ACM computing surveys (CSUR)},
  volume={37},
  number={4},
  pages={316--344},
  year={2005},
  publisher={ACM New York, NY, USA}
}

@article{geppert2022trusted,
  title={Trusted execution environments: Applications and organizational challenges},
  author={Geppert, Tim and Deml, Stefan and Sturzenegger, David and Ebert, Nico},
  journal={Frontiers in Computer Science},
  volume={4},
  pages={930741},
  year={2022},
  publisher={Frontiers Media SA}
}

@inproceedings{wang2024understanding,
  title={Understanding Ethereum Mempool Security under Asymmetric $\{$DoS$\}$ by Symbolized Stateful Fuzzing},
  author={Wang, Yibo and Tang, Yuzhe and Li, Kai and Ding, Wanning and Yang, Zhihua},
  booktitle={33rd USENIX Security Symposium (USENIX Security 24)},
  pages={4747--4764},
  year={2024}
}

@inproceedings{myakala2025intent,
  title={Intent-Driven Decentralization: Architectures for Private, Scalable, and Autonomous Systems},
  author={Myakala, Praveen Kumar and Jonnalagadda, Anil Kumar and Thomas, Sooraj George},
  booktitle={2025 Global Conference in Emerging Technology (GINOTECH)},
  pages={1--11},
  year={2025},
  organization={IEEE}
}

@inproceedings{wang2024achieving,
  title={Achieving Privacy-Preserving Optimizer Architecture for Intent Execution on EVM Blockchain},
  author={Wang, Weijie and Deng, Haotian and Liang, Jinwen and Zhang, Chuan},
  booktitle={Proceedings of the 6th ACM International Symposium on Blockchain and Secure Critical Infrastructure},
  pages={1--9},
  year={2024}
}

@article{jauernig2020trusted,
  title={Trusted execution environments: properties, applications, and challenges},
  author={Jauernig, Patrick and Sadeghi, Ahmad-Reza and Stapf, Emmanuel},
  journal={IEEE Security \& Privacy},
  volume={18},
  number={2},
  pages={56--60},
  year={2020},
  publisher={IEEE}
}

@article{zheng2021survey,
  title={A survey of Intel SGX and its applications},
  author={Zheng, Wei and Wu, Ying and Wu, Xiaoxue and Feng, Chen and Sui, Yulei and Luo, Xiapu and Zhou, Yajin},
  journal={Frontiers of Computer Science},
  volume={15},
  number={3},
  pages={153808},
  year={2021},
  publisher={Springer}
}

@article{pinto2019demystifying,
  title={Demystifying arm trustzone: A comprehensive survey},
  author={Pinto, Sandro and Santos, Nuno},
  journal={ACM computing surveys (CSUR)},
  volume={51},
  number={6},
  pages={1--36},
  year={2019},
  publisher={ACM New York, NY, USA}
}

@inproceedings{li2022systematic,
  title={A systematic look at ciphertext side channels on AMD SEV-SNP},
  author={Li, Mengyuan and Wilke, Luca and Wichelmann, Jan and Eisenbarth, Thomas and Teodorescu, Radu and Zhang, Yinqian},
  booktitle={2022 IEEE Symposium on Security and Privacy (SP)},
  pages={337--351},
  year={2022},
  organization={IEEE}
}

@inproceedings{orenbach2017eleos,
  title={Eleos: ExitLess OS services for SGX enclaves},
  author={Orenbach, Meni and Lifshits, Pavel and Minkin, Marina and Silberstein, Mark},
  booktitle={Proceedings of the Twelfth European Conference on Computer Systems},
  pages={238--253},
  year={2017}
}

@inproceedings{paju2023sok,
  title={Sok: A systematic review of tee usage for developing trusted applications},
  author={Paju, Arttu and Javed, Muhammad Owais and Nurmi, Juha and Savim{\"a}ki, Juha and McGillion, Brian and Brumley, Billy Bob},
  booktitle={Proceedings of the 18th International Conference on Availability, Reliability and Security},
  pages={1--15},
  year={2023}
}

@article{hardjono2019decentralized,
  title={Decentralized trusted computing base for blockchain infrastructure security},
  author={Hardjono, Thomas and Smith, Ned},
  journal={Frontiers in Blockchain},
  volume={2},
  pages={24},
  year={2019},
  publisher={Frontiers Media SA}
}

@article{de2021analysis,
  title={An analysis of the fees and pending time correlation in Ethereum},
  author={de Azevedo Sousa, Jos{\'e} Eduardo and Oliveira, Vin{\'\i}cius and Valadares, J{\'u}lia and Dias Goncalves, Glauber and Moraes Villela, Saulo and Soares Bernardino, Heder and Borges Vieira, Alex},
  journal={International Journal of Network Management},
  volume={31},
  number={3},
  pages={e2113},
  year={2021},
  publisher={Wiley Online Library}
}

@article{reepu2025stablecoins,
  title={Stablecoins: The Cornerstone of DeFi},
  author={Reepu and Kumar, Pawan and Taneja, Sanjay and Ozen, Ercan},
  journal={Decentralized Finance and the End of Traditional Banking},
  pages={165--177},
  year={2025},
  publisher={Wiley Online Library}
}

@article{wang2021non,
  title={Non-fungible token (NFT): Overview, evaluation, opportunities and challenges},
  author={Wang, Qin and Li, Rujia and Wang, Qi and Chen, Shiping},
  journal={arXiv preprint arXiv:2105.07447},
  year={2021}
}

@article{castro2021financial,
  title={Financial intermediation and risk in decentralized lending protocols},
  author={Castro-Iragorri, Carlos and Ramirez, Julian and Velez, Sebastian},
  journal={arXiv preprint arXiv:2107.14678},
  year={2021}
}

@inproceedings{chemaya2024power,
  title={The power of default: Measuring the effect of slippage tolerance in decentralized exchanges},
  author={Chemaya, Nir and Liu, Dingyue and McLaughlin, Robert and Ruaro, Nicola and Kruegel, Christopher and Vigna, Giovanni},
  booktitle={International Conference on Financial Cryptography and Data Security},
  pages={192--208},
  year={2024},
  organization={Springer}
}

@article{meister2024gas,
  title={Gas fees on the Ethereum blockchain: from foundations to derivative valuations},
  author={Meister, Bernhard K and Price, Henry CW},
  journal={Frontiers in Blockchain},
  volume={7},
  pages={1462666},
  year={2024},
  publisher={Frontiers Media SA}
}

@inproceedings{white2022characterizing,
  title={Characterizing the OpenSea NFT marketplace},
  author={White, Bryan and Mahanti, Aniket and Passi, Kalpdrum},
  booktitle={Companion Proceedings of the Web Conference 2022},
  pages={488--496},
  year={2022}
}

@inproceedings{mizrahi2021optimizing,
  title={Optimizing Merkle proof size for blockchain transactions},
  author={Mizrahi, Avi and Koren, Noam and Rottenstreich, Ori},
  booktitle={2021 International Conference on COMmunication Systems \& NETworkS (COMSNETS)},
  pages={299--307},
  year={2021},
  organization={IEEE}
}

@inproceedings{wan2019evaluating,
  title={Evaluating the impact of network latency on the safety of blockchain transactions},
  author={Wan, Luming and Eyers, David and Zhang, Haibo},
  booktitle={2019 IEEE International Conference on Blockchain (Blockchain)},
  pages={194--201},
  year={2019},
  organization={IEEE}
}

@article{wilbaut2008survey,
  title={A survey of effective heuristics and their application to a variety of knapsack problems},
  author={Wilbaut, Christophe and Hanafi, Said and Salhi, Said},
  journal={IMA journal of management Mathematics},
  volume={19},
  number={3},
  pages={227--244},
  year={2008},
  publisher={Oxford University Press}
}

@inproceedings{rouhani2017performance,
  title={Performance analysis of ethereum transactions in private blockchain},
  author={Rouhani, Sara and Deters, Ralph},
  booktitle={2017 8th IEEE international conference on software engineering and service science (ICSESS)},
  pages={70--74},
  year={2017},
  organization={IEEE}
}

@article{rani2023academic,
  title={Academic payment tokenization: an online payment system for academia utilizing non-fungible tokens and permissionless blockchain},
  author={Rani, Prity and Sachan, Rohit Kumar and Kukreja, Sonal},
  journal={Procedia Computer Science},
  volume={230},
  pages={347--356},
  year={2023},
  publisher={Elsevier}
}

@article{munk2022names,
  title={When Are Names Similar Or the Same? Introducing the Code Names Matcher Library},
  author={Munk, Moshe and Feitelson, Dror G},
  journal={arXiv preprint arXiv:2209.03198},
  year={2022}
}

@article{kim2024optimal,
  title={Optimal gas fee minimization in defi: Enhancing efficiency and security on the ethereum blockchain},
  author={Kim, Heesang and Kim, Dohoon},
  journal={IEEE Access},
  year={2024},
  publisher={IEEE}
}

\appendix
\section{Syntax Definition of ICL}
\label{syntax_definition}

\begin{lstlisting}[basicstyle=\ttfamily\small,
  basicstyle=\ttfamily\footnotesize,
  keywordstyle=\color{blue},
  commentstyle=\color{green!50!black},
  stringstyle=\color{orange},
  numbersep=5pt,
  showstringspaces=false,
  tabsize=2,
  breaklines=true,
  breakatwhitespace=false,
  language=C,
  frame=single]
grammar IntentDSL;
program: triggerStatement+;

LPARENT : '('; RPARENT : ')'; LBRACK : '['; RBRACK : ']';
LOGIC_AND : 'and'; LOGIC_OR : 'or'; LOGIC_NOT : 'not';
EQ : '=='; NEQ : '!='; LT : '<'; GT : '>'; LE : '<='; GE : '>=';
ADD : '+'; SUB : '-'; MUL : '*'; DIV : '/'; MOD : '%';
BALANCE : 'balance'; PRICE : 'price'; SLIPPAGE : 'slippage'; FEE : 'fee'; WALLET : 'wallet';
USDT : 'USDT'; USDC : 'USDC'; ETH : 'ETH'; DAI : 'DAI'; BTC : 'BTC'; WBTC : 'WBTC'; WETH : 'WETH'; UNI : 'UNI'; SUSHI : 'SUSHI'; AAVE_token : 'AAVE'; MATIC : 'MATIC'; COMP : 'COMAP';
AAVE : 'Aave'; UNISWAP : 'Uniswap'; COMPOUND : 'Compound'; YEARN : 'Yearn'; SUSHISWAP : 'Sushiswap'; CURVE : 'Curve'; ONEINCH : '1inch'; POLYGON : 'Polygon'; AVAX : 'Avax';

DEC_INT : '0' | ([1-9][0-9]*);
DEC_FLOAT
    : [0-9]*'.'[0-9]*(('p'|'P'|'e'|'E')('+'|'-')?[0-9]+)?
    | [0-9]+[.]?[0-9]*('p'|'P'|'e'|'E')(('+'|'-')?[0-9]+)?;
PRIVATE_KEY : [A-Fa-f0-9]+; KEY : '0'('x'|'X')[0-9A-Fa-f]+;
TIME : [0-9][0-9][0-9][0-9] '-' ('0'[1-9] | '1'[0-2]) '-' ('0'[1-9] | [12][0-9] | '3'[01]) 'T' ([01]?[0-9] | '2'[0-3]) ':' [0-5]?[0-9] ':' [0-5]?[0-9];
SEMI : ';'; LINE_COMMENT : '//' .*? '\r'?'\n' -> skip; COMMENT : '/*' .*? '*/' -> skip; BLANK : [ \t\r\n]+ -> skip;

condition : orExpression;
timeCondition : 'time before' TIME | 'time after' TIME | 'time during' TIME 'to' TIME;
orExpression : andExpression (LOGIC_OR andExpression)*;
andExpression : (comparisonExpression | timeCondition)(LOGIC_AND (comparisonExpression| timeCondition))*;
comparisonExpression : comparisonElement comparisonOperator comparisonElement;
comparisonElement : walletBalance | assetPrice | number asset | number | SLIPPAGE | FEE | LPARENT orExpression RPARENT;
binaryExpression : lowBinaryExpression (highBinaryOperator lowBinaryExpression)*;
lowBinaryExpression : unaryExpression (lowBinaryOperator unaryExpression)*;
unaryExpression : (unaryOperator)* primaryExpression;
primaryExpression : number | LPARENT (binaryExpression | unaryExpression) RPARENT;
logicalOperator : LOGIC_AND | LOGIC_OR;
comparisonOperator : EQ | NEQ | LT | GT | LE | GE;
highBinaryOperator : MUL | DIV | MOD;
lowBinaryOperator : ADD | SUB;
unaryOperator : ADD | SUB | LOGIC_NOT;
number : DEC_INT | DEC_FLOAT;

walletBalance : BALANCE wallet;
assetPrice : PRICE asset;
amount : binaryExpression asset;
asset : USDT | USDC | ETH | DAI | BTC | WBTC | UNI | SUSHI | AAVE_token | MATIC | COMP | WETH;
wallet : WALLET LBRACK KEY RBRACK;
platform : AAVE | UNISWAP | COMPOUND | YEARN | SUSHISWAP | CURVE | ONEINCH | POLYGON | AVAX;

triggerStatement : ('trigger' condition 'then')? statement ('checking' condition)? SEMI;

statement : transferStatement | borrowStatement | repayBorrowStatement | swapStatement | addLiquidityStatement | removeLiquidityStatement | stakeStatement | buyNFTStatement | sellNFTStatement | simpleStakeStatement | simpleBuyNFTStatement | simpleSellNFTStatement;

transferStatement : 'transfer' amount 'from' wallet 'to' wallet;
borrowStatement : 'borrow' amount 'for' wallet 'from' platform;
repayBorrowStatement : 'repay' amount 'from' wallet 'to' platform;
swapStatement : 'swap' amount 'from' wallet 'for' asset 'on' platform;
addLiquidityStatement : 'add' amount ',' amount 'to' platform 'receiving liquidity token to' wallet;
removeLiquidityStatement : 'remove' amount ',' amount 'from' platform 'returning' DEC_INT 'liquidity' ('of token' LBRACK KEY RBRACK)? 'from' wallet;
simpleStakeStatement : 'stake' amount 'from' wallet 'on' platform;
simpleBuyNFTStatement : 'buy NFT' LBRACK KEY RBRACK 'in collection' LBRACK KEY RBRACK 'using at most' amount 'from' wallet;
simpleSellNFTStatement : 'sell NFT' LBRACK KEY RBRACK 'in collection' LBRACK KEY RBRACK 'from' wallet 'for at least' amount;

stakeStatement : 'stake' amount 'from' wallet (stakeStrategy)?;
stakeStrategy : 'using' (stakeStrategyQualifiers)* 'strategy';
stakeStrategyQualifiers : 'low-risk' | 'middle-risk' | 'high-risk' | 'short-term' | 'middle-term' | 'long-term';

buyNFTStatement : 'buy' (NFTQualifiers)* 'NFT using at most' amount 'from' wallet;
NFTQualifiers : 'mainstream' | 'popular' | 'rare' | 'inexpensive' | 'price-increasing' | 'price-decreaseing';
NFTPlatform : 'OpenSea' | 'Rarible' | 'SuperRare' | 'Foundation' | 'Mintable' | 'BakerySwap' | 'LooksRare';

sellNFTStatement : 'sell NFT' LBRACK KEY RBRACK 'in collection' LBRACK KEY RBRACK 'from' wallet (sellNFTStartegy)?;
sellNFTStartegy : 'using' (sellNFTStrategyQualifiers)* 'strategy';
sellNFTStrategyQualifiers : 'time-saving' | 'profitable';
\end{lstlisting}

\section{Multiple Knapsack for Minimizing Dependencies}
\label{multiple_knapsack}

\begin{algorithm}[htbp]
    \footnotesize
	\DontPrintSemicolon
	\SetKwInOut{Input}{Input}
	\SetKwInOut{Output}{Output}
	\SetKw{Continue}{continue}
    
    \caption{Multiple Knapsack} 
        \Input{
    		 $packages$, $capacities$
    	} 
    	
    	\Output{
    		List of selected packages
    	}

        \tcp{capacities: Backpack capacity (item type $\rightarrow$ item capacity)}
        \tcp{packages: List of item quantities per package (package $\rightarrow$ item type $\rightarrow$ item quantity)}
        \tcp{DP map, record backpack status and its corresponding maximum number of packages}
        
        $dp \leftarrow Map(String, BigInteger)$
        
        $selectedPackages \leftarrow Map(String, List)$
        
        $initCapacities \leftarrow Map(String, BigInteger)$

        \tcp{Initialization: When capacity is at maximum, no package is selected}
        \For{\textbf{each} $itemType$ \textbf{in} $capacities$}
        {
            $initCapacities[itemType] \leftarrow 0$
        }

        $dp[createKey(initCapacities)] \leftarrow 0$
        
        $selectedPackages[createKey(initCapacities)] \leftarrow []$

        \For{\textbf{each} $packageItem$ \textbf{in} $packages$}
        {
            $packageContents \leftarrow packageItem.getAssetIncrease()$
            
            $currentStates \leftarrow list\ of\ keys\ from\ dp$

            \For{\textbf{each} $stateKey$ \textbf{in} $currentStates$}
            {
               $state \leftarrow parseKey(stateKey)$, $newState \leftarrow state$, $valid \leftarrow \textbf{true}$

                \For{\textbf{each} $(itemType, itemCount)$ \textbf{in} $packageContents$}
                {
                    $currentCapacity \leftarrow newState[itemType]$
                    
                    $newCapacity \leftarrow currentCapacity + itemCount$
                    
                    \If{$newCapacity > capacities[itemType]$}
                    {
                        $valid \leftarrow \textbf{false}$, \textbf{break}
                    }
                    
                    $newState[itemType] \leftarrow newCapacity$
                }

                \If{$valid$}
                {
                    $newStateKey \leftarrow createKey(newState)$, $newValue \leftarrow dp[stateKey] + 1$
                    
                    \If{$newStateKey$ \textbf{not in} $dp$ \textbf{or} $newValue > 
                    dp[newStateKey]$}
                    {
                        $dp[newStateKey] \leftarrow newValue$
                        
                        $newPackageList \leftarrow selectedPackages[stateKey] + packageItem$
                        
                        $selectedPackages[newStateKey] \leftarrow newPackageList$
                    }
                }
            }
        }
        
        $bestStateKey \leftarrow key\;of\;the\;entry\;in\;dp\;with\;the\;maximum\;value$
        
        \Return $selectedPackages[bestStateKey]$
\end{algorithm}
\section{Execution Optimizer Theoretical Derivations}
\label{appendix:optimizer_theory}

\subsection{Predicted Balance and Safety Invariant}

For each transaction $tx_j$, let $\Delta^+_j(w,a)$ and $\Delta^-_j(w,a)$ denote the increase and consumption of asset $a$ in wallet $w$. With initial balance $B_0(w,a)$, the predicted balance before $x_j$ executes is
\[
\hat{B}_j(w, a) = B_0(w, a) + \sum_{tx_i \in \text{Dom}(tx_j)} \left[ \Delta^+_i(w,a) - \Delta^-_i(w,a) \right],
\]
where $\text{Dom}(tx_j)$ are the ancestors of $tx_j$ in the dependency graph.  
The safety condition requires:
\[
\forall (w,a): \quad \hat{B}_j(w,a) \geq \Delta^-_j(w,a).
\]

\subsection{Redundant Dependency Elimination}

To maximize concurrency, we seek a minimal parent set $S_j \subseteq \text{Pa}(tx_j)$ such that the invariant holds:
\[
\min_{S_j \subseteq \text{Pa}(tx_j)} |S_j| \quad \text{s.t.} \quad \hat{B}_j^{(S_j)}(w,a) \geq \Delta^-_j(w,a) \ \forall (w,a).
\]

This corresponds to a multi-dimensional bounded subset-sum (knapsack) problem, where each parent provides a “budget” of asset effects. We solve this using a dynamic programming–based pruning algorithm, as integrated in Phase~2 of Algorithm~\ref{optimally_execute_tx_set}. After pruning, the DAG is updated to preserve transitive dependencies (see Fig.~\ref{update_the_DAG}).

\subsection{Complexity Analysis}

The worst-case complexity is $O(2^m)$ where $m$ is the number of direct parents of a node. In practice, transaction dependency graphs are sparse, so exact solutions are tractable. For larger graphs, approximate or heuristic solvers~\cite{wilbaut2008survey} can be used to balance accuracy and efficiency.

\section{Prompts}
\label{prompts}

\paragraph{Attempt to fulfill the transaction intents in the dataset using the designed ICL.}

You are a decentralized finance (DeFi) trading assistant. I have designed a decentralized fi-nance trading tool in the form of a domain-specific language (DSL). Below is the grammar definition of this language: 

<grammar>  

You will receive a trading intent for a decentralized finance transaction. Try to accomplish this intent using the appropriate instructions from the language.  

If the intent can be fulfilled, output the corresponding instruction directly (without any extra content).  

If the intent cannot be fulfilled, output 'DISABLED'.  

Here is the trading intent to fulfill: <intent> 

\paragraph{Attempt to implement the transaction intents using a specific programming language.}

You are a decentralized finance intent translator. You need to translate the following intent into the shortest possible specified code (or code framework). 

Just output the code text without comments or any other content. Now please translate the following intent into <language> : <intent>

\end{document}